\documentclass[pra,%
 reprint,
preprintnumbers,
nofootinbib,
 amsmath,amssymb,
 aps,
]{revtex4-2}

\usepackage{graphicx}
\usepackage{dcolumn}
\usepackage{bm}
\usepackage{cancel}
\usepackage{ulem}
\usepackage[colorlinks=true,
            linkcolor=blue,
            citecolor=blue,
            urlcolor=blue]{hyperref}

\usepackage[dvipsnames]{xcolor}

\begin{document}

\preprint{OU-HET 1285}
\preprint{KEK-TH-2755}
\preprint{KU-PH-039}

\title{Classification of Higgs sectors from group theoretical properties of UV gauge theories}

\author{Shinya Kanemura}
\email{kanemu@het.phys.sci.osaka-u.ac.jp}
\affiliation{Department of Physics, The University of Osaka, Toyonaka, Osaka 560-0043, Japan}

\author{Yushi Mura}
\email{yushi37@post.kek.jp}
\affiliation{KEK Theory Center, Tsukuba, Ibaraki 305–0801, Japan}%

\author{Tetsuo Shindou}
\email{shindou@cc.kogakuin.ac.jp}
\affiliation{Division of Liberal-Arts, Kogakuin University,
Hachioji, Tokyo, 192-0015, Japan}%


\begin{abstract}
Extended Higgs sectors are often introduced to explain phenomena beyond the standard model (BSM).
The existence of multiple scalar fields may cause the Landau pole below the Planck scale.
In this case, the low-energy theory may be replaced by an asymptotic-free gauge theory.
In this paper, we consider an $\mathrm{SU}(2)$ gauge theory with confinement as such an ultraviolet theory of the extended Higgs sectors.
We investigate the relation between scalar particle contents at the low energy and group theoretical properties of fundamental fermions of the gauge theory.
We find that particle contents of various extended Higgs sectors previously proposed to explain the BSM problems are deduced by each charge assignment of flavor symmetry of the fundamental fermions of the $\mathrm{SU}(2)$ gauge symmetry.
Our findings may provide a new picture for the ultraviolet completion of the extended Higgs sectors. 
\end{abstract}

\maketitle


\section{\label{sec:Intro} Introduction}

Although the Higgs boson was discovered at the Large Hadron Collider~\cite{ATLAS:2012yve,CMS:2012qbp}, the true shape of the Higgs sector, the number of the scalar fields, their representations under the gauge symmetries, dynamics behind, and so on, remain unknown.
While the discovered Higgs boson is consistent with the one in the standard model (SM) under the theoretical and experimental uncertainties~\cite{ATLAS:2022vkf,CMS:2022dwd}, there is no theoretical guiding principle for the Higgs sector of the SM. 
Therefore, the possibility of non-standard Higgs sectors has been discussed in various physics contexts.
In fact, there remain phenomena which the SM cannot explain, such as tiny neutrino masses, dark matter, and baryon asymmetry of the Universe.
Some of them might be related to physics of non-standard (extended) Higgs sectors.

From experiments, several strong constraints on the shape of the Higgs sector are indicated, such that the electroweak rho parameter~\cite{Veltman:1977kh,Chanowitz:1978uj,Einhorn:1981cy} is very close to unity~\cite{Baak:2012kk,Baak:2014ora,ParticleDataGroup:2024cfk}, and that the flavor changing neutral currents are strongly suppressed~\cite{UTfit:2006onp,UTfit:2007eik,Isidori:2010kg}.
The LHC data~\cite{ATLAS:2022vkf,CMS:2022dwd} also indicate that an extended Higgs sector, if it exists, has to show the SM-like behavior at the electroweak scale at least approximately.
Future experiments may detect evidence of the extended Higgs sectors.

Extended Higgs models can explain the tiny neutrino masses by the type-II seesaw mechanism~\cite{Magg:1980ut,Cheng:1980qt,Lazarides:1980nt,Mohapatra:1980yp} or by quantum effects~\cite{Zee:1980ai,Wolfenstein:1981kw,Zee:1985id,Babu:1988ki}.
By imposing an unbroken $Z_2$ symmetry to non-minimal Higgs sectors, weakly interacting dark matter candidates can be provided~\cite{Deshpande:1977rw}.
Tiny neutrino masses and dark matter simultaneously are explained by the combination of these mechanisms in various models~\cite{Tao:1996vb,Ma:2006km,Gustafsson:2012vj,Krauss:2002px}.
Extended Higgs sectors can also make the electroweak phase transition strongly first order which cannot be realized in the SM Higgs sector~\cite{Anderson:1991zb,Funakubo:1993jg,Espinosa:1993bs,Davies:1994id,Cline:1996mga,Huber:2000mg,Barger:2008jx,Chiang:2014hia,Patel:2012pi}.
As extended Higgs sectors naturally contain additional CP violating phases, the scenario of electroweak baryogenesis~\cite{Kuzmin:1985mm} is possible.
Many models with extended Higgs sectors have been considered for electroweak baryogenesis~\cite{Cline:2021iff,Espinosa:2011eu,Cline:2012hg,Turok:1990zg,Cline:1995dg,Fromme:2006cm,Cline:2011mm,Basler:2021kgq,Enomoto:2021dkl,Enomoto:2022rrl,Kanemura:2023juv,Idegawa:2023bkh,Aiko:2025tbk,Inoue:2015pza}.
Furthermore, combination of these ideas leads to $\mathrm{TeV}$ scale models which can explain all the three problems simultaneously~\cite{Aoki:2008av,Aoki:2022bkg,Enomoto:2024jyc}.

Each model mentioned above has specific particle contents for the Higgs sector.
While some of such models are phenomenologically successful, a specific pattern of additional scalar fields is introduced in an ad-hoc way.
However, such a specific scalar sector should be deduced as a low-energy effective theory of a more fundamental theory at high energies.

In the SM with the $125~\mathrm{GeV}$ Higgs boson, the quartic coupling of the Higgs boson falls down to be negative below the Planck scale due to the large mass of the top quark, and the vacuum becomes meta-stable~\cite{Bezrukov:2012sa,Degrassi:2012ry,Buttazzo:2013uya,Bednyakov:2015sca}.
On the contrary, in the extended Higgs sectors, the ultraviolet (UV) behavior of the couplings can be changed by quantum effects of additional scalar bosons~\cite{Inoue:1982ej,Flores:1982pr,Nie:1998yn,Ferreira:2009jb,Kominis:1993zc,Kanemura:1999xf,Chowdhury:2015yja,Goudelis:2013uca,Kanemura:2023wap,Bandyopadhyay:2025ilx,Blasi:2017xmc}.
In general, at the one-loop level, additional scalar bosons give positive contributions to the beta functions of the scalar quartic couplings.
If the contributions are large, then the running coupling constants may blow up at the Landau pole below the Planck scale.
Such extended Higgs models should then have a cutoff scale, above which they are replaced by a more fundamental theory.

In this paper, we consider a possibility of an asymptotic-free gauge theory with confinement as a candidate of the fundamental theory above the cutoff scale.
We here adopt $\mathrm{SU}(2)$ (hereafter we denote $\mathrm{SU}(2)_H$) as such a gauge symmetry as the simplest example~\cite{Lewis:2011zb,Kanemura:2012uy,Cacciapaglia:2014uja,Arbey:2015exa,Arthur:2016dir,Arthur:2016ozw,Drach:2017btk}.
We assume that the scalar bosons in the extended Higgs sectors appear at the low energy as hadron-like bound states of the fundamental fermions due to the strong interaction of $\mathrm{SU}(2)_H$.
Particle contents and their symmetry properties in the extended Higgs sectors will be inherited from the nature of flavor symmetry of the fundamental fermions.

Many scenarios, in which the electroweak symmetry breaking sector is a low-energy effective theory of a fundamental theory above the cutoff scale, have been discussed in different contexts, such as technicolor models~\cite{Weinberg:1975gm,Susskind:1978ms,Kaplan:1991dc,Farhi:1980xs,Hill:2002ap}, composite Higgs models~\cite{Kaplan:1983fs,Kaplan:1983sm,Agashe:2004rs,Mrazek:2011iu}, models for confinement with supersymmetry~\cite{Harnik:2003rs,Kanemura:2012uy}, and so on.

In our scenario, the SM gauge group $\mathrm{SU}(2)_L \otimes \mathrm{U}(1)_Y$ is embedded into the flavor symmetry of the $\mathrm{SU}(2)_H$ gauge theory.
Unlike the technicolor theory, spontaneous symmetry breaking (SSB) of the SM gauge group is assumed to be caused by the scalar bosons within the IR effective theory.
We do not specify those scalar bosons as the Nambu--Goldstone (NG) bosons of SSB of the flavor symmetries. 
As a consequence of the confinement of the $\mathrm{SU}(2)_H$ gauge theory, the extended Higgs sectors for various phenomenological models are classified by the number of the fundamental fermions and their nature under the flavor symmetry and the SM gauge group.

This paper is organized as follows.
In Sec.~\ref{sec:SU2theory}, we give general discussions on the $\mathrm{SU}(2)_H$ gauge theory.
In Sec.~\ref{sec:Higgsmodels}, we classify extended Higgs models in terms of the flavor number and the charge assignment in the $\mathrm{SU}(2)_H$ gauge theory.
The discussions and the conclusions are given in Sec.~\ref{sec:Discussions} and Sec.~\ref{sec:Conclusions}.

\section{$\mathrm{SU}(2)_H$ gauge theory \label{sec:SU2theory}}

In this section, we discuss the $\mathrm{SU}(2)_H$ gauge theory with $n$ fundamental fermions.
The theory is asymptotic-free or conformal, when $n \lesssim 10$~\cite{Karavirta:2011zg,Neil:2011yag}.
We assume that, at the energy scale below a scale $\Lambda$, the fermions are confined due to the strong interaction of $\mathrm{SU}(2)_H$.
It is known that $n \lesssim 4.7$ is needed for the massless fermions to be confined~\cite{Appelquist:1999hr,Neil:2011yag}.

\subsection{Flavor symmetry}
We first discuss properties of a fundamental fermion $\psi$ under $\mathrm{SU}(2)_H$ and its Dirac conjugate $\overline{\psi}$.
They are written by 
\begin{align}
  \psi_{h} = 
  \begin{pmatrix}
    \chi_{\alpha h} \\ \eta^{\dot{\alpha}}_{h}
  \end{pmatrix},~~ 
  \overline{\psi}^h = \big(\eta^{* \alpha h}, \chi^{* h}_{\dot{\alpha}} \big),
  \label{eq:dirac fermion}
\end{align}
where $\chi_\alpha$ and $\eta^{* \alpha}$ ($\chi^*_{\dot{\alpha}}$ and $\eta^{\dot{\alpha}}$) are the $\bm{2}$ ($\bm{2}^*$) spinors of $\mathrm{SL}(2,\mathbb{C})$.
The spinor index is raised or lowered by the anti-symmetric tensors $\epsilon^{\alpha \beta} = \epsilon^{\dot{\alpha} \dot{\beta}} = \epsilon_{\alpha \beta} = \epsilon_{\dot{\alpha} \dot{\beta}}$ ($\epsilon^{12} = +1$).
The subscript (superscript) $h$ represents $\bm{2}$ ($\bm{2}^*$) of $\mathrm{SU}(2)_H$, and they can be connected by the anti-symmetric tensors $\epsilon_{H h_1 h_2} = \epsilon_H^{h_1 h_2}$ ($\epsilon_{H 1 2} = +1$). 
A proper charge conjugation for $\psi$ is defined as 
\begin{align}
  \tilde{\psi}_{h_1} \equiv \epsilon_{H h_1 h_2} (\psi^C)^{h_2}  =
  \begin{pmatrix}
    \eta^*_{\alpha h_1} \\ \chi^{* \dot{\alpha}}_{h_1}
  \end{pmatrix}.
\end{align} 
It is notable that, for any index $I = \{\alpha,\dot{\alpha},h \}$, $a^{I} b_{I} = - a_{I} b^{I} $ is satisfied.

Let us see the flavor symmetry in the kinetic term of the $\mathrm{SU}(2)_H$ sector.
With $n$ flavor fermions $\psi_i$ ($i=1,...,n$), it is given by
\begin{align}
  \mathcal{L}_{\mathrm{kin}} = \overline{\psi_i} i \gamma^\mu D_\mu \psi_i,
\end{align}
where $D_\mu$ is the covariant derivative of the $\mathrm{SU}(2)_H$ gauge field, and the summation of the flavor index $i$ is implicit.
In terms of $\chi$ and $\eta$, it is subsequently rewritten by
\begin{align}
  \mathcal{L}_{\mathrm{kin}}&= \chi^{*h_1}_{i \dot{\alpha}} (i \overline{\sigma}^\mu)^{\dot{\alpha} \beta} (D_\mu)^{h_2}_{h_1} \chi_{i \beta h_2} +\eta^{* \alpha h_1}_{i} (i\sigma^\mu)_{\alpha \dot{\beta}} (D_\mu)^{h_2}_{h_1} \eta_{i h_2}^{\dot{\beta}} \notag \\
  &=(\overline{\bm{T}})^{h_1}_{\dot{\alpha}} (i \overline{\sigma}^\mu)^{\dot{\alpha} \beta} (D_\mu)^{h_2}_{h_1} (\bm{T})_{\alpha h_2},
\end{align}
where the sign conventions for the Pauli matrices $\sigma^\mu = (\mathbb{I}_{2},-\sigma^a)$ and $\overline{\sigma}^\mu = (\mathbb{I}_{2},\sigma^a)$ $(a = 1,2,3)$ are taken.
We have defined the flavor multiplet $\bm{T}$ as 
\begin{align}
  \bm{T} \equiv (\chi_1,\eta^*_1,...,\chi_n, \eta_n^* )^\intercal \equiv (T_1,...,T_n)^\intercal,
\end{align} 
In this kinetic term, there is $\mathrm{SU}(2 n)$ symmetry.

Without loss of generality, the flavor index can be taken for the basis where the Dirac mass is diagonalized: the mass term of the $\mathrm{SU}(2)_H$ sector is, in general, given by 
\begin{align}
  \mathcal{L}_{\mathrm{mass}} &= -m_i \overline{\psi_i} \psi_i -\mathcal{M}_{ij} \Big( \overline{\tilde{\psi}_i} \psi_j -\overline{\psi_i} \tilde{\psi}_j \Big) \notag \\
  &=-\frac{1}{2} (\bm{T}^{\intercal})^{\alpha h}
  \bm{m}
  (\bm{T})_{\alpha h} 
  -\frac{1}{2} (\bm{T}^{\intercal})^{\alpha h}
  \bm{M}
  (\bm{T})_{\alpha h} + \mathrm{h.c.},
\end{align}
where $m_i$ and $\mathcal{M}_{ij}$ are the Dirac and Majorana masses, respectively.
We assume all of the masses are real.
Defining $M_{ij} \equiv \mathcal{M}_{ij} - \mathcal{M}_{ji}$, the mass matrices are given by 
\begin{align}
  \bm{m} &= \bm{m}_{n\times n} \otimes -i \sigma_2, \notag \\ 
  \bm{M} &=
  \bm{M}_{n\times n} \otimes \mathbb{I}_2,
\end{align}
where 
\begin{align}
  &\bm{m}_{n\times n} = \mathrm{diag}(m_1,...,m_n), \notag \\
  &\bm{M}_{n\times n} =
\begin{pmatrix}
  0         & M_{12}    & \cdots & \cdots    & M_{1n} \\
  -M_{12}   & 0         & \cdots & \cdots    & M_{2n} \\
  \vdots    & \vdots    & \ddots &           & \vdots \\
  \vdots    & \vdots    &        & 0         & M_{\,n-1\,n} \\
  -M_{1n}   & -M_{2n}   & \cdots & -M_{\,n-1\,n} & 0
\end{pmatrix}.
\end{align}

Depending on the form of the mass term, various explicit breakings of the $\mathrm{SU}(2n)$ flavor symmetry are realized.
For example, when we have $m_i = m$ and $M_{ij} = 0$, the explicit symmetry breaking is $\mathrm{SU}(2n) \to \mathrm{Sp}(2n)$.
When we have $M_{ij} = 0$ $(j=1,...,n)$ for the $i$-th fermion $T_i$, we can see an $\mathrm{SU}(2)$ invariance $T_i \to U T_i$ ($U \in \mathrm{SU}(2)_i$).
Therefore, if $M_{ij} = 0$ for all $i$ and $j$, we have the $\mathrm{SU}(2)^n (= \otimes^n_{i=1} \mathrm{SU}(2)_i)$ symmetry in this Lagrangian.
We note that, the $\mathrm{U}(1)_i$ symmetries are realized as the subgroup of the $\mathrm{SU}(2)_i$ symmetries, under which the flavor multiplets transform as $T_i \to \exp(i \sigma_3 \theta_i) T_i$.
This symmetry corresponds to the global $\mathrm{U}(1)_V$ symmetry for $\psi_i$; the Lagrangian is invariant under $\psi_i \to \exp(i\theta_i) \psi_i$.
By contrast, the $\mathrm{U}(1)_A$ symmetries ($\psi_i \to \exp(i\theta_i \gamma^5) \psi_i$) are explicitly broken by the Dirac mass terms.

When $\chi_i = \eta^*_i$ is satisfied, the corresponding $i$-th fermion $\psi_i$ satisfies the Majorana condition, $\psi_i = \tilde{\psi}_i$.
In this paper, we do not discuss cases where such a condition is imposed.

\subsection{Discrete symmetries \label{sec:discrete}}

First, we define the parity transformation for the multiplets $\bm{T}$, which are used to classify bound states.
The usual parity transformation defined in terms of $\psi_i$ deduces
\begin{align}
  \chi_{i \alpha h} (x) \xrightarrow{P} \eta^{\dot{\alpha}}_{i h} (x_P), ~~~~
  \eta^{\dot{\alpha}}_{i h} (x) \xrightarrow{P} \chi_{i \alpha h} (x_P),
\end{align}
where $x = (t,\bm{x})$ and $x_P = (t,-\bm{x})$.
However, this transformation is not invariant under the flavor symmetry.
Following Ref.~\cite{Drach:2017btk}, we define a \textit{proper} parity transformation $Q$ as 
\begin{align}
  \chi_{i \alpha h}(x)  \xrightarrow{Q} \beta \eta^{\dot{\alpha}}_{i h} (x_P) , ~~~~
  \eta^{\dot{\alpha}}_{i h} (x)  \xrightarrow{Q} \beta \chi_{i \alpha h} (x_P) ,
  \label{eq:Qtrans}
\end{align}
where $\beta$ is an arbitrary phase.

In terms of $\bm{T}$, the transformation is given by 
\begin{align}
  (\bm{T})_{\alpha h_1}^{A} \xrightarrow{Q}~ 
  Q \epsilon^{\dot{\alpha} \dot{\beta}} \epsilon_{H h_1 h_2} (\bm{T}^*)^{\overline{A} h_2}_{\dot{\beta}} = Q (\bm{T}^*)^{\overline{A} \dot{\alpha}}_{h_1},
\end{align}
where 
\begin{align}
    Q = \mathbb{I}_n \otimes \begin{pmatrix}
        0 & \beta \\ \beta^* & 0
    \end{pmatrix}.
\end{align}
We here have omitted the space-time variables.
The index of the (complex conjugate) representation of the flavor symmetry is denoted by $A$ ($\overline{A}$).
Since the multiplets transform under the flavor symmetry as $\bm{T}^{A} \to U^{AB} \bm{T}^{B}$ or $\bm{T}^{* \overline{A}} \to U^{\overline{A} \overline{B}} \bm{T}^{* \overline{B}}$, we need to satisfy the conditions,
\begin{align}
  Q U^* = U Q,
\end{align}
so that $\mathcal{Q} \bm{T} \mathcal{Q}^{-1}$ transforms the same as $\bm{T}$.
We cannot satisfy these conditions if the flavor symmetry is maximal, i.e. $U \in \mathrm{SU}(2n)$.
However, in the case of $\mathrm{Sp}(2n)$, the solution is $\beta = \pm i$, by taking which $Q$ is proportional to
\begin{align}
  J^{A \overline{B}} = (\mathbb{I}_n \otimes i\sigma_2)^{A \overline{B}}.
\end{align}
For any of $U \in \mathrm{Sp}(2n)$, $U^\intercal J U = J$ is satisfied.
In the following discussions, we consider the explicit breaking of $\mathrm{SU}(2n)$ by the mass term, and thus the maximal symmetry is $\mathrm{Sp}(2n)$.
Therefore, taking $\beta = +i$ is enough to discuss the bound states.
As a result, the parity transformation is given by
\begin{align}
  (\bm{T})^A_{\alpha h} &\xrightarrow{Q} iJ^{A \overline{B}} (\bm{T}^{*})^{\overline{B} \dot{\alpha}}_h = i (\bm{T}^{*})^{A \dot{\alpha}}_h. 
\end{align}

For later discussions, we here consider a discrete symmetry $Z_N = \otimes^n_{i=1} Z_N^i$ of the multiplet $\bm{T}$.
We consider the $Z_N$ transformation commuting to the transformations under the remaining flavor symmetry and the parity.
The $Z_N$ transformation is defined as 
\begin{align}
  \bm{T} \xrightarrow{Z_n}
  \begin{pmatrix}
    \omega_1 & & & & \\
    & \omega_1^* & & & \\
    & & \ddots & & \\
    & & & \omega_n & \\
    & & & & \omega_n^*
  \end{pmatrix} \bm{T} \equiv Z \bm{T},
\end{align}
where $\omega_i \equiv \exp(2 \pi i /N_i) ~ (N_i \in \mathbb{Z})$.
The Dirac mass term is invariant under this transformation.

When the $\mathrm{Sp}(2n)$ symmetry exists in the Lagrangian, it must be satisfied that
\begin{align}
  Z U = U Z,
\end{align}
where $U \in \mathrm{Sp}(2n)$.
From Schur's Lemma, all of $\omega_i$ must have the same $Z_2$ charge as $\omega_i = \omega = \pm 1$.
When the $\mathrm{SU}(2)^n$ symmetry is present, the $Z_2$ charges for each flavor can be different: i.e. $\omega_i = \omega_i^* = \pm 1$ is allowed.
By taking $\omega_i$ in these ways, the matrix can be written as $Z = \mathrm{diag}(\omega_1, ..., \omega_n) \otimes \mathbb{I}_2$, which commutes with the parity transformation matrix $Q = i J = i (\mathbb{I}_n \otimes i \sigma_2)$.

\subsection{Bound states}

The scalar bosons, which are bound states of two fermions, are classified by irreducible representations of the flavor symmetry in the UV Lagrangian.
Letting $\bm{T}^A$ be an irrep. of a flavor symmetry, the scalar bound states are either 
\begin{align}
  (\bm{T})^{A \alpha h} (\bm{T})^B_{\alpha h}~~~\text{or}~~~(\bm{T}^*)^{A \dot{\alpha} h} (\bm{T}^*)^B_{\dot{\alpha} h}.
\end{align}
We here have assumed that the irrep. is real or pseudo-real.
From the discussions given in Sec.~\ref{sec:discrete}, it is shown that the states written by 
\begin{align}
  S^{AB}_{\pm} = (\bm{T})^{A \alpha h} (\bm{T})^B_{\alpha h} \pm (\bm{T}^*)^{A \dot{\alpha} h} (\bm{T}^*)^B_{\dot{\alpha} h},
\end{align}
are parity even and odd for $+$ and $-$, respectively.

We note that totally symmetric representations with respect to the flavor indices cannot be the scalar bosons due to the anti-commuting property of the fermions.
Such a symmetric representation can appear as the vector-like boson, e.g. 
\begin{align}
  V^{\mu A B} = (\bm{T})^{A \alpha h} \sigma^\mu_{\alpha \dot{\beta}} (\bm{T}^*)^{B \dot{\beta}}_h + (A \leftrightarrow B).
\end{align}

\section{Extended Higgs models \label{sec:Higgsmodels}}

In this section, we discuss extended Higgs models deduced by the $\mathrm{SU}(2)_H$ theory with $n$ flavor fermions $\psi_i$ ($i = 1,...,n$).
In our setup, the SM quarks and leptons are singlets under $\mathrm{SU}(2)_H$, and the scalar sectors of various extended Higgs models are realized by the confinement of the fundamental fermions of $\mathrm{SU}(2)_H$.

As mentioned in the introduction, we do not specify the scalar bosons as the NG boson, and therefore we do not focus on the spontaneous breaking of the flavor symmetry induced by the confinement\footnote{
For the case of $n=2$, lattice simulations have shown that spontaneous breaking $\mathrm{SU}(4)\to \mathrm{Sp}(4)$ indeed occurs~\cite{Lewis:2011zb,Arthur:2016dir,Arthur:2016ozw,Drach:2017btk}. 
}.
The chiral symmetry $\mathrm{SU}(2n)$ is explicitly broken by the fermion mass terms.
The SM gauge group is embedded in the flavor symmetry of the $\mathrm{SU}(2)_H$ Lagrangian, which is assumed to be spontaneously broken in the context of extended Higgs models around the electroweak scale $v \simeq 246~\mathrm{GeV}$.

Our focus is not on the concrete mass spectrum in the extended Higgs models, but rather on revealing how those models are classified into the flavor symmetry of the UV Lagrangian and the charge assignment of the SM gauge group.

In the next subsection, we present examples of the extended Higgs models derived from the $n$ flavor $\mathrm{SU}(2)_H$ gauge theory, for $n$ up to $4$.
We then discuss the Yukawa interaction in the extended Higgs models.  

\subsection{$n=1$ case}
We suppose the $\mathrm{SU}(2)_H$ theory with a fermion $T_1 = (\chi_1, \eta_1^*)^\intercal$.
The flavor symmetry is $\mathrm{Sp}(2) \simeq \mathrm{SU}(2)$ regardless of the Dirac mass $m_1$.
This $\mathrm{SU}(2)$ can be regarded as the $\mathrm{SU}(2)_L$ gauge symmetry by introducing the minimal interaction~\cite{Mrazek:2011iu}
\begin{align}
    \mathcal{L} = (\overline{T_1})^{h}_{\dot{\alpha}} (i \overline{\sigma}^\mu)^{\dot{\alpha} \beta} \Big(-i g \frac{\bm{\sigma}}{2} \cdot \bm{W}_\mu \Big) (T_1)_{\beta h},
\end{align}
where $\bm{W}_\mu = (W_\mu^1,W_\mu^2,W_\mu^3)^\intercal$ and $g$ are the $\mathrm{SU}(2)_L$ gauge bosons and coupling, respectively.
However, the possible representation of the scalar is only the singlet state of $\mathrm{SU}(2)_L$.
In addition, the $\mathrm{U}(1)_Y$ interaction cannot be embedded into the flavor symmetry.
Therefore, the SM Higgs doublet is not realized in this case.

\subsection{$n=2$ case}

The smallest flavor number including the SM Higgs is $n=2$.
The $\mathrm{SU}(2)_H$ sector is composed of two fermions, $T_1 = (\chi_1,\eta_1^*)^\intercal$ and $T_2 = (\chi_2,\eta_2^*)^\intercal$.
The mass parameters are $m_1,m_2$, and $M_{12}$.

\subsubsection*{Class~1: Real singlet extension of the SM}

We first consider the case that the Majorana mass is zero, $M_{12} = 0$.
In the $\mathrm{SU}(2)_H$ Lagrangian, in general, we have $\mathrm{SU}(2)_1 \otimes \mathrm{SU}(2)_2$ symmetry for $T_1$ and $T_2$, respectively.
If $m_1 = m_2$, the symmetry is expanded into $\mathrm{Sp}(4)$.

We regard $\mathrm{SU}(2)_1$ and $\mathrm{U}(1)_2 (\subset \mathrm{SU}(2)_2)$ as the gauged $\mathrm{SU}(2)_L$ and $\mathrm{U}(1)_Y$ symmetries.
The $\mathrm{U}(1)_Y$ gauge interaction with the gauge coupling $g^\prime$ and the hypercharge $Y=y_2$ is given by 
\begin{align}
   \mathcal{L} = (\overline{T_2})_{\dot{\alpha}}^h (i\overline{\sigma}^\mu)^{\dot{\alpha} \beta} \Big(-i g^\prime 2y_2 \frac{\sigma_3}{2} B_\mu \Big) (T_2)_{\beta h}.
   \label{eq:T2U1Yint}
\end{align}
To realize the SM Higgs doublet, we set $y_2 = 1/2$.

The latter $\mathrm{SU}(2)_2$ can be regarded as a global $\mathrm{SU}(2)_R$ symmetry of the SM.
This $\mathrm{SU}(2)_R$ symmetry guarantees the rho parameter to be unity at the tree level~\cite{Sikivie:1980hm}, which is explicitly broken by the $\mathrm{U}(1)_Y$ interaction and the Yukawa interaction, which will be explained in Sec.~\ref{sec:Yukawa}.

By denoting $a_1$ and $b_1$ ($a_2$ and $b_2$) as the indices of $\mathrm{SU}(2)_L$ ($\mathrm{SU}(2)_R$), the scalar bosons are given by 
\begin{align}
  \phi_1^{\pm} \equiv S_{a_1 b_1}^{\pm} \epsilon_L^{a_1 b_1},~~ \phi_2^{\pm} \equiv S_{a_2 b_2}^{\pm}\epsilon_R^{a_2 b_2},~~\mathbb{M}_{a_1 a_2}^{\pm} \equiv S_{a_1 a_2}^{\pm},
\end{align}
where $\epsilon_L$ and $\epsilon_R$ are the anti-symmetric tensors for each $\mathrm{SU}(2)_L$ and $\mathrm{SU}(2)_R$, respectively, and the same notation as that for $\mathrm{SU}(2)_H$ is used.
The sign represents the eigenvalue of the parity transformation.

The former two are the singlet states under $\mathrm{SU}(2)_L \otimes \mathrm{SU}(2)_R$.
In addition, they are not charged by $\mathrm{U}(1)_Y$.
By taking the Hermitian conjugate, it is shown that
\begin{align}
  (\phi_i^\pm)^\dagger = \mp \phi_i^\pm,~~(i = 1,2)
\end{align}
which indicates that $i \phi^+_i$ and $\phi^-_i$ are parity-even and parity-odd hermitian scalar fields, respectively.
In general, these fields are independent, since the $\mathrm{U}(1)_A$ transformation acting on $\psi_i$, which can be regarded as an $\mathrm{SO}(2)$ transformation on $(i\phi^+_i,\phi^-_i)^\intercal$, is explicitly broken by $m_1$ and $m_2$.

The last one is the bi-doublet $(\bm{2},\bm{2})$ state of two $\mathrm{SU}(2)$.
For the bi-doublet, 
\begin{align}
  (\mathbb{M}^{\pm}_{a_1 a_2})^\dagger = \pm \mathbb{M}^{\pm}_{b_1 b_2} \epsilon^{b_1 a_1}_L \epsilon^{b_2 a_2}_R = \pm \mathbb{M}^{\pm a_1 a_2},
\end{align}
is shown.
Therefore, the pairs of $(a_1,a_2) = (1,1),(2,2)$ and $(1,2),(2,1)$ components are connected by the Hermitian conjugate.
When we define 
\begin{align}
  \mathbb{M}_{a_1 a_2} \equiv -(\mathbb{M}^+_{a_1 a_2} + i \mathbb{M}^-_{a_1 a_2}),
  \label{eq:bidoublet}
\end{align}
each component of the matrix can be identified with the complex scalars due to the $\mathrm{U}(1)_i \subset \mathrm{SU}(2)_i$ symmetry.
We then obtain the $(\bm{2},\bm{2}^*)$ matrix corresponding to $\mathbb{M}_{\mathrm{SM}} = (i \sigma_2 \Phi^*_{\mathrm{SM}}, \Phi_{\mathrm{SM}}) = (\tilde{\Phi}_{\mathrm{SM}}, \Phi_{\mathrm{SM}})$ as 
\begin{align}
  \mathbb{M}_{\mathrm{SM}} \propto \mathbb{M}_{a_1}^{a_2} = 
  \begin{pmatrix}
    \mathbb{M}^+_{12} + i \mathbb{M}^-_{12} & -(\mathbb{M}^+_{11} + i\mathbb{M}^-_{11}) \\
    (\mathbb{M}^+_{11} + i\mathbb{M}^-_{11})^\dagger & (\mathbb{M}^+_{12} + i \mathbb{M}^-_{12})^\dagger
  \end{pmatrix}_{a_1}^{a_2}.
\end{align}

\begin{table}[t]
\setlength{\tabcolsep}{8pt}
\centering
\begin{tabular}{|c|c|c|c|}
\hline
 & $\mathbb{M}$ & $\phi_1^\pm$ & $\phi_2^\pm$  \\ \hline
$Y$ & $1/2$ & $0$ & $0$ \\ \hline
$L$ & $\bm{2}$ & $\bm{1}$ & $\bm{1}$  \\ \hline
$R$ & $\bm{2}$ & $\bm{1}$ & $\bm{1}$ \\  \hline
\end{tabular}
\caption{
Scalar particle contents below $\Lambda$ in the $n=2$ case.
The value of the hypercharge is the highest weight of $\mathrm{SU}(2)_R$.
}
\label{tab:class1}
\end{table}

The particle contents in Class~1 are summarized in Table~\ref{tab:class1}.
This class leads to the real singlet extension of the SM (xSM)~\cite{Hill:1987ea,Krasnikov:1997nh,OConnell:2006rsp}.
In this table, we have omitted the flavor indices, and the value of the hypercharge is the highest weight of $\mathrm{SU}(2)_R$.

In the case of $m_1 = m_2$ in the limit of neglecting the breaking effects of $\mathrm{Sp}(4)$, $\mathbb{M}_{a_1 a_2}$ and $\phi_i^\pm$ are unified into the anti-symmetric representations $\bm{5}+\bm{1}$.

We note that, if the Majorana mass $M_{12}$ is non-zero but $m_1 = m_2 = 0$, there is an $\mathrm{SU}(2)_1 \otimes \mathrm{SU}(2)_2$ symmetry for the two multiplets $(\chi_1, \chi_2)^\intercal$ and $(\eta^*_1, \eta^*_2)$.
By replacing the label of the particles as $\eta^*_1 \leftrightarrow  \chi_2$, we obtain the same $\mathrm{SU}(2)_H$ Lagrangian as that with $m_1 = m_2$ and $M_{12} = 0$.
Therefore, the scalar bound states are $\bm{5}+\bm{1}$ of $\mathrm{Sp(4)}$.

\subsection{$n=3$ case \label{sec:n=3case}}
In this subsection, we show the extended Higgs models deduced by the $\mathrm{SU}(2)_H$ theory with the three flavor fermions $T_1,T_2$, and $T_3$.
In the $n=3$ case, the rich flavor symmetry allows for various extended Higgs models, including the 2HDM as well as models addressing dark matter and neutrino masses.

We consider a case where $m_1, m_2, m_3 \neq 0$ and $M_{ij} = 0$ are taken.
Following similar arguments as in the $n=2$ case, we have the $\mathrm{SU}(2)_1 \otimes \mathrm{SU}(2)_2 \otimes  \mathrm{SU}(2)_3$ symmetry for $T_1, T_2$ and $T_3$, respectively.

In order to embed the SM gauge group, $\mathrm{SU}(2)_1$ ($\mathrm{SU}(2)_L$) and the third component of $\mathrm{SU}(2)_2$ ($\mathrm{SU}(2)_{R_1}$) are gauged in the same way as in the $n=2$ case, i.e., with $y_2 = 1/2$. 
The remaining $\mathrm{SU}(2)_3$ ($\mathrm{SU}(2)_{R_2}$) can likewise be gauged by assigning the hypercharge $Y = y_3$ to $T_3$.
We assign a $Z_2$ charge $\omega_3 = \pm 1$ for $T_3$.

\begin{table}[t]
\setlength{\tabcolsep}{6pt}
\centering
\begin{tabular}{|c|c|c|c|c|c|c|c|}
\hline
 & $\mathbb{M}_{12}$ & $\mathbb{M}_{13}$ & $\mathbb{M}_{23}$ & $\phi_1^\pm$ & $\phi_2^\pm$ & $\phi_3^\pm$  \\ \hline
 $Z_2$ & $+$ & $\omega_3$ & $\omega_3$ & $+$ & $+$ & $+$ \\ \hline
 $\spadesuit$ $Y$ & $1/2$ & $y_3$ & $1/2 + y_3$ & $0$ & $0$ & $0$ \\ \hline
 $\spadesuit$ $L$ & $\bm{2}$ & $\bm{2}$ & $\bm{1}$ & $\bm{1}$ & $\bm{1}$ & $\bm{1}$  \\ \hline
 $\diamondsuit$ $R_1$ & $\bm{2}$ & $\bm{1}$ & $\bm{2}$ & $\bm{1}$ & $\bm{1}$ & $\bm{1}$ \\  \hline
 $\heartsuit$ $R_2$ & $\bm{1}$ & $\bm{2}$ & $\bm{2}$ & $\bm{1}$ & $\bm{1}$ & $\bm{1}$ \\ \hline
\end{tabular}
\caption{
Scalar particle contents below $\Lambda$ in the $n=3$ case.
In Class~2 (Class~3), $y_3 = 1/2$ and $\omega_3 = + 1$ ($\omega_3 = -1$) are taken.
In Class~4, $y_3 = 3/2$ and $\omega_3 = + 1$ are taken.
The hypercharge values correspond to the highest weights of $\mathrm{SU}(2)_{R_1} \otimes \mathrm{SU}(2)_{R_2}$.
}
\label{tab:class2}
\end{table}

With a similar discussion for the $n=2$ case, which we do not repeat here, we obtain scalar bound states that are eigenstates of the parity.
There are three bi-doublet fields and three singlet states under the flavor symmetry.
The $\mathrm{SU}(2)_L \otimes \mathrm{SU}(2)_{R_1}$ bi-doublet $\mathbb{M}_{12}$, which is composed of $T_1$ and $T_2$, corresponds to the SM Higgs bi-doublet $\mathbb{M}_{\mathrm{SM}}$.
The $\mathbb{M}_{13}$ ($\mathbb{M}_{23}$) field is the $\mathrm{SU}(2)_L \otimes \mathrm{SU}(2)_{R_2}$ ($\mathrm{SU}(2)_{R_1} \otimes \mathrm{SU}(2)_{R_2}$) bi-doublet, which is the tensor product of $T_1$ and $T_3$ ($T_2$ and $T_3$).
The $\phi_i^\pm$ $(i=1,2,3)$ fields are totally anti-symmetric singlet representations of each $\mathrm{SU}(2)$.
Their hypercharge is determined by $Y= (T_{R_1})_3+2y_3(T_{R_2})_3$, where $(T_{R_1})_3$ and $(T_{R_2})_3$ denote the representation matrices of the third components of $\mathrm{SU}(2)_{R_1}$ and $\mathrm{SU}(2)_{R_2}$, respectively.

A summary of the particle contents in the $n=3$ case is presented in Table~\ref{tab:class2}.
As before, flavor indices are omitted.
The hypercharge values correspond to the highest weights of $\mathrm{SU}(2)_{R_1} \otimes \mathrm{SU}(2)_{R_2}$.

The symmetries marked by the spade suit $\spadesuit$ have to be always exact.
As we have noted, since the $\mathrm{U}(1)_Y$ interactions violate $\mathrm{SU}(2)_{R_1}$, it is an approximate symmetry (marked by diamond $\diamondsuit$).
On the contrary, when we set $y_3 = 0$, the $\mathrm{SU}(2)_{R_2}$ symmetry can be exact.
When we have a non-zero hypercharge of $T_3$ or other interactions breaking $\mathrm{SU}(2)_{R_2}$ (e.g. the Yukawa interaction), it becomes not exact.
This kind of symmetry, which can be either exact or not exact, is marked by the heart suit $\heartsuit$.

Up to now, the charge assignment of $y_3$ and $\omega_3$ has been left unspecified.
In the following three classes of the charge assignment, we find that the $n=3$ case effectively encompasses the two Higgs doublet model (2HDM)~\cite{Lee:1973iz,Branco:2011iw}, the Inert doublet model (IDM)~\cite{Deshpande:1977rw}, the Tao--Ma model~\cite{Tao:1996vb,Ma:2006km}, the complex singlet extension of the SM (CxSM)~\cite{Barger:2008jx}, the scalar singlet dark matter model (SDM)~\cite{Silveira:1985rk,McDonald:1993ex,Burgess:2000yq}, the Zee model~\cite{Zee:1980ai,Wolfenstein:1981kw}, and the Zee--Babu model~\cite{Zee:1985id,Babu:1988ki}.

\subsubsection*{Class~2: 2HDM, Zee model, and CxSM}

As a charge assignment, we take $y_3 = 1/2$ and $\omega_3 = +1$.
In this class, $\mathbb{M}_{12}$ and $\mathbb{M}_{13}$ form part of the 2HDM scalar potential, which is composed of two $\mathrm{SU}(2)_L$ doublets $\Phi_1$ and $\Phi_2$ with $Y=1/2$. 
Making the $\mathrm{SU}(2)_{R_1}$ and $\mathrm{SU}(2)_{R_2}$ indices raised, we denote $(\bm{2},\bm{2}^*,\bm{1})$ and 
$(\bm{2},\bm{1},\bm{2}^*)$ as $\mathbb{M}_1 = (\tilde{\Phi}_1, \Phi_1)$ and $\mathbb{M}_2 = (\tilde{\Phi}_2, \Phi_2)$, respectively.

The transformation law under the three $\mathrm{SU}(2)$ is given by
\begin{align}
    \mathbb{M}_{1} \to L \mathbb{M}_{1} R_1^\dagger, ~~~ \mathbb{M}_{2} \to L \mathbb{M}_{2} R_2^\dagger,
\end{align}
where $L \in \mathrm{SU}(2)_L$, $R_1 \in \mathrm{SU}(2)_{R_1}$, and $R_{2} \in \mathrm{SU}(2)_{R_2}$.
The $\mathrm{U}(1)_Y$ transformation is given by 
\begin{align}
    \mathbb{M}_{1,2} \to \mathbb{M}_{1,2} \exp\Big(-ig^\prime \frac{\sigma^3}{2} \Big).
\end{align}
There are four invariants under $\mathrm{SU}(2)_L \otimes \mathrm{U}(1)_Y$,
\begin{align}
    \mathrm{Tr}[\mathbb{M}_{1}^\dagger \mathbb{M}_1], ~ \mathrm{Tr}[\mathbb{M}_{2}^\dagger \mathbb{M}_2],~\mathrm{Tr}[\mathbb{M}_{1}^\dagger \mathbb{M}_2], ~ \mathrm{Tr}[\mathbb{M}_{1}^\dagger \mathbb{M}_2 \sigma_3], 
\end{align}
and the former two are also invariant under $\mathrm{SU}(2)_{R_1} \otimes \mathrm{SU}(2)_{R_2}$.
Therefore, if we assume that couplings violating $\mathrm{SU}(2)_{R_1}$ or $\mathrm{SU}(2)_{R_2}$ are only $g^\prime$ or the Yukawa couplings, the potential terms written by the latter two invariants are expected to be small~\cite{Mrazek:2011iu}.

More concretely, the potential which is invariant under $\mathrm{SU}(2)_{R_1} \otimes \mathrm{SU}(2)_{R_2}$ is written up to dimension four as 
\begin{align}
  V^{(d \le 4)}_{R_1,R_2} (\Phi_1,\Phi_2) = &~\mu_1^2 \Phi_1^\dagger \Phi_1 + \mu_2^2 \Phi_2^\dagger \Phi_2 + \frac{\lambda_1}{2} |\Phi_1|^4 \notag \\
  &+ \frac{\lambda_2}{2} |\Phi_2|^4 + \lambda_3 |\Phi_1|^2 |\Phi_2|^2.
  \label{eq:2HDMpotential}
\end{align}
The potential terms violating $\mathrm{SU}(2)_{R_1} \otimes \mathrm{SU}(2)_{R_2}$ written by
\begin{align}
  &V^{(d \le 4)}_{\cancel{R}_1 \cancel{R}_2}(\Phi_1,\Phi_2) = \mu_3^2 \Big(\Phi_1^\dagger \Phi_2 + \mathrm{h.c.} \Big) + \lambda_4 |\Phi_1^\dagger \Phi_2|^2 \notag \\
  &+ \Big(\frac{\lambda_5}{2} (\Phi_1^\dagger \Phi_2) + \lambda_6 |\Phi_1|^2 + \lambda_7 |\Phi_2|^2 \Big) (\Phi_1^\dagger \Phi_2) + \mathrm{h.c.},
  \label{eq:2HDMpotential_v}
\end{align}
should be suppressed by the effects violating $\mathrm{SU}(2)_{R_1} \otimes \mathrm{SU}(2)_{R_2}$.
We here assume all couplings are real.
We also have higher dimensional operators in the potential, which are suppressed by the cutoff scale ($\sim \Lambda$) of this model.
Since the significant one-loop corrections to the rho parameter in the 2HDM come from effects violating the custodial symmetry, this potential does not cause such a large correction.
If only $\Phi_1$ acquires the vacuum expectation value, while $\Phi_2$ does not, or if the condition $\lambda_1 \simeq \lambda_2 \simeq \lambda_3$ is also satisfied\footnote{
This parametrization can be realized as a consequence of an $\mathrm{SO}(5)$ symmetry in the potential~\cite{Pilaftsis:2011ed,BhupalDev:2014bir,Pilaftsis:2016erj,Darvishi:2019ltl,Darvishi:2021txa}.
}, the mixing among the neutral scalar bosons is also suppressed~\cite{BhupalDev:2014bir}, which satisfies the current LHC measurements~\cite{ATLAS:2022vkf,CMS:2022dwd}.
This class of the 2HDM has been discussed in Refs.~\cite{Pilaftsis:2011ed,Pilaftsis:2016erj}.

If we discard the $\mathrm{SU}(2)_{R_1} \otimes \mathrm{SU}(2)_{R_2}$ symmetry by introducing a non-zero Majorana mass $M_{23}$, those symmetry breaking terms in the potential can be large\footnote{
In this case, the kinetic mixing terms
\begin{align}
    \mathrm{Tr}[(D_\mu \mathbb{M}_1)^\dagger D^\mu \mathbb{M}_2 ], ~~ \mathrm{Tr}[(D_\mu \mathbb{M}_1)^\dagger  D^\mu \mathbb{M}_2 \sigma_3],
\end{align}
which are invariant under $\mathrm{SU}(2)_L \otimes \mathrm{U}(1)_Y$ are induced.
By using $\mathrm{GL}(2,\mathbb{C}) / \mathrm{U}(2)$ transformation with respect to $\Phi_1$ and $\Phi_2$, the canonical basis can be taken~\cite{Haber:2006ue}.
}.
However, in such a case, significant deviations of the rho parameter and the Higgs couplings from the SM prediction are expected.

Since this class also includes the $\mathrm{SU}(2)_L$ singlets with $Y=1$ in $\mathbb{M}_{23}$, the particle contents of the original Zee model are contained.
When we denote the charged singlet scalar by $h^\pm$, the relevant Lagrangian of the Zee-model is given by
\begin{align}
    -\mathcal{L}_{\text{Zee}} = f_L \overline{\ell_L^C} \ell_L h^+ + \mu \tilde{\Phi}_1^\dagger \Phi_2 h^- + \mathrm{h.c.}, 
    \label{eq:ZeeYukawa}
\end{align}
where $\ell_L$ ($e_R$) is the left-handed (right-handed) SM lepton doublet (singlet), and $f_L$ is anti-symmetric with respect to interchanging the lepton flavors.
In the Zee-model, the active neutrino masses are generated at the one-loop level as
\begin{align}
    m_\nu \sim  \frac{f_L \zeta \mu m_{\ell}^2}{(16 \pi^2) M^2} \mathcal{I}_{\text{Zee}},
\end{align}
where $\mathcal{I}$ is a dimensionless factor as a function of masses, and we denote a representative scale of new physics by $M (< \Lambda)$.
The Yukawa coupling associated with $\Phi_2$ is denoted by $\zeta$.
We note that the three-point scalar coupling $\mu$ should be suppressed due to the $\mathrm{SU}(2)_{R_1} \otimes \mathrm{SU}(2)_{R_2}$ symmetry.
Since it is known that the observed data cannot be explained by the original Zee model~\cite{Frampton:1999yn,Koide:2001xy}, some extensions have been discussed~\cite{Balaji:2001ex,Frampton:2001eu}.

In addition, there is the $\mathrm{SU}(2)_L$ singlet with $Y=0$ contained in $\mathbb{M}_{23}$.
Although this field has vanishing hypercharge, it can be identified with a complex scalar field if the global $\mathrm{U}(1)_{3}$ symmetry is exact.
As long as such breaking effects vanish or can be neglected, this class effectively includes the CxSM.

\subsubsection*{Class~3: IDM, Tao--Ma model, and SDM}

By taking $y_3 = 1/2$ and $\omega_3 = -1$, the bi-doublets $\mathbb{M}_{13}$ and $\mathbb{M}_{23}$ can be $Z_2$-odd.
The $Z_2$ parity of $T_3$ is the only difference between Class~2 and Class~3. 
In this class, the shape of the potential made by the two $Y=1/2$ bi-doublets $\mathbb{M}_1$ ($Z_2$-even) and $\mathbb{M}_2$ ($Z_2$-odd) is similar to Eqs.~\eqref{eq:2HDMpotential}-\eqref{eq:2HDMpotential_v}.
However, $\mu_3^2,\lambda_6$ and $\lambda_7$ terms are not allowed due to the exact $Z_2$ symmetry.
The $\lambda_4$ and $\lambda_5$ terms are allowed but should be suppressed due to the approximate $\mathrm{SU}(2)_{R_1} \otimes \mathrm{SU}(2)_{R_2}$ symmetry.
This potential is that of the IDM.
In this model, the $Z_2$-even $\mathbb{M}_1$ should be regarded as the SM Higgs bi-doublet, and the $Z_2$-odd $\mathbb{M}_2$ includes the scalar dark matter candidates.

When we additionally have $Z_2$-odd right handed neutrinos (RHNs) $N$ with the Majorana masses $M_{N}$, the particle contents include that of the Tao-Ma model.
The relevant Lagrangian is given by
\begin{align}
    -\mathcal{L}_{\text{Tao--Ma}} = \frac{\lambda_5}{2} (\Phi_1^\dagger \Phi_2)^2 + h \overline{\ell_L} \Phi_2 N + \mathrm{h.c.}, 
\end{align}
and the neutrino masses are given as
\begin{align}
    m_\nu \sim \frac{h^\dagger h \lambda_5 v^2}{(16 \pi^2) M_N} \mathcal{I}_{\text{Tao--Ma}}.
    \label{eq:TaoMa}
\end{align}

We note that, in this class, the $Z_2$-odd $\mathrm{SU}(2)_L$ singlet scalar with $Y=0$ is also contained in $\mathbb{M}_{23}$. 
Therefore, this class also includes the SDM, in which the $Z_2$-odd singlet can serve as another dark matter candidate as the singlet scalar dark matter.

\subsubsection*{Class~4: Zee--Babu model}

By taking $y_3 = 3/2$ and $\omega_3 = +1$, in addition to the SM Higgs doublet, we have an additional $\mathrm{SU}(2)_L$ doublet with $Y=3/2$ in $\mathbb{M}_{13}$.
We also have $Y=1$ and $Y=2$ $\mathrm{SU}(2)_L$ singlets in $\mathbb{M}_{23}$, which are denoted by $h^+$ and $k^{++}$, respectively.

This class realizes the particle contents of the Zee-Babu model, in which the neutrino masses are generated at the two-loop level.
As in the Zee model, the relevant Lagrangian of the Zee-Babu model is given by 
\begin{align}
    -\mathcal{L}_{\text{Zee--Babu}} &= f_L \overline{\ell_L^C} \ell_L h^+ + f_R \overline{e_R^C} e_R k^{++} + \mathrm{h.c.} \notag \\
    &+ \sigma h^- h^- k^{++} + \mathrm{h.c.},
\end{align}
where $f_R$ is symmetric under changing lepton flavor indices.
The $\sigma$ term should be suppressed by the $\mathrm{SU}(2)_{R_1} \otimes \mathrm{SU}(2)_{R_2}$ symmetry.
The neutrino masses are represented by 
\begin{align}
    m_\nu \sim  \frac{f_L^\intercal f_R f_L \sigma m_{\ell} m_{{\ell^\prime}}}{(16 \pi^2)^2 M^2} \mathcal{I}_{\text{Zee--Babu}}.
\end{align}

\subsection{$n=4$ case}
In this subsection, we discuss the $n=4$ case, in which $T_1,T_2,T_3$, and $T_4$ are involved.
In the following, we consider two ways to embed the $\mathrm{SU}(2)_L$ gauge symmetry.
The first one is a straightforward extension of the previous $n=3$ case, where $\mathrm{SU}(2)_L$ is embedded in $\mathrm{SU}(2)$ of $T_1$.
In the other way, it is embedded into a subgroup of $\mathrm{Sp}(6)$ in $(T_1,T_2,T_3)^\intercal$.

The former way realizes several models generating the neutrino masses at the loop level, such as the Aoki--Kanemura--Yagyu (AKY) model~\cite{Aoki:2011yk}, the Gustafsson--No--Rivera (GNR) model~\cite{Gustafsson:2012vj}, the Krauss--Nasri--Trodden (KNT) model~\cite{Krauss:2002px}, and the Aoki--Kanemura--Seto (AKS) model~\cite{Aoki:2008av}.
The next-to-minimal 2HDM (N2HDM)~\cite{Chen:2013jvg} is also included. 
In the latter way, the $\mathrm{SU}(2)_L$ triplet model, which is represented by the Georgi--Machacek (GM) model~\cite{Georgi:1985nv,Chanowitz:1985ug}, is realized.

\subsection*{Models for the neutrino mass}

\begin{table*}[t]
\setlength{\tabcolsep}{6pt}
\centering
\begin{tabular}{|c|c|c|c|c|c|c|c|c|c|c|c|}
\hline
 & $\mathbb{M}_{12}$ & $\mathbb{M}_{13}$ & $\mathbb{M}_{14}$ & $\mathbb{M}_{23}$ & $\mathbb{M}_{24}$ & $\mathbb{M}_{34}$ & $\phi_1^\pm$ & $\phi_2^\pm$ & $\phi_3^\pm$ & $\phi_4^\pm$  \\ \hline
 $\spadesuit$ $Z_2$ & $+$ & $\omega_3$ & $\omega_4$ & $\omega_3$ & $\omega_4$ & $\omega_3 \omega_4$ & $+$ & $+$ & $+$ & $+$ \\ \hline
 $\spadesuit$ $Y$ & $1/2$ & $ y_3$ & $ y_4$ & $1/2 + y_3$ & $1/2 + y_4$ & $y_3 + y_4$ & $0$ & $0$ & $0$ & $0$ \\ \hline
 $\spadesuit$ $L$ & $\bm{2}$ & $\bm{2}$ & $\bm{2}$ & $\bm{1}$ & $\bm{1}$ & $\bm{1}$ & $\bm{1}$ & $\bm{1}$ & $\bm{1}$ & $\bm{1}$  \\ \hline
 $\diamondsuit$ $R_1$ & $\bm{2}$ & $\bm{1}$ & $\bm{1}$ & $\bm{2}$ & $\bm{2}$ & $\bm{1}$ & $\bm{1}$ & $\bm{1}$ & $\bm{1}$ & $\bm{1}$  \\  \hline
 $\heartsuit$ $R_2$ & $\bm{1}$ & $\bm{2}$ & $\bm{1}$ & $\bm{2}$ & $\bm{1}$ & $\bm{2}$ & $\bm{1}$ & $\bm{1}$ & $\bm{1}$ & $\bm{1}$  \\ \hline
 $\heartsuit$ $R_3$ & $\bm{1}$ & $\bm{1}$ & $\bm{2}$ & $\bm{1}$ & $\bm{2}$ & $\bm{2}$ & $\bm{1}$ & $\bm{1}$ & $\bm{1}$ & $\bm{1}$  \\ \hline
\end{tabular}
\caption{
Scalar particle contents below $\Lambda$ in the $n=4$ case for models generating the neutrino masses.
In Class 5 and Class 6, the charge assignments $(y_3, \omega_3, y_4, \omega_4)= (1/2, -1, 3/2, -1)$ and $ (1/2, +1, 1/2, -1)$ are taken, respectively.
The hypercharge values shown here are the highest weights of $\mathrm{SU}(2)_{R_1} \otimes \mathrm{SU}(2)_{R_2} \otimes \mathrm{SU}(2)_{R_3}$.
}
\label{tab:n=4_1}
\end{table*}

In addition to the $n=3$ case, we introduce $T_4$, which has the hypercharge $y_4$ and the $Z_2$ charge $\omega_4$.
By assuming that different Dirac masses $m_1,m_2,m_3$, and $m_4$ and $M_{ij} = 0$, we additionally have $\mathrm{SU}(2)_{R_3}$ in the $T_4$ sector.
There are new bi-doublet and singlet fields, $\mathbb{M}_{14}, \mathbb{M}_{24}$, $\mathbb{M}_{34}$, and $\phi_4^\pm$.
The summary of the particle contents is given in Table~\ref{tab:n=4_1}.

\subsubsection*{Class~5: AKY model, GNR model, and KNT model}

In this class, we assign the hypercharge and the $Z_2$ charge as $y_3 = 1/2$, $y_4 = 3/2$, and $\omega_3 = \omega_4 = -1$.
In addition to the SM Higgs doublet ($\Phi_1 \in \mathbb{M}_{12}$), we have the $Z_2$-odd additional $\mathrm{SU}(2)_L$ doublets with $Y=1/2$ ($\Phi_2 \in \mathbb{M}_{13}$) and $Y=3/2$ ($\Phi_{3/2} \in \mathbb{M}_{14}$).

When a $Z_2$-odd vector-like fermion $\psi$ with the Dirac mass $m_\psi$ is added in this class, the AKY model, where the neutrino masses are generated at the one-loop level, is realized\footnote{
To reproduce the experimental data for the neutrino, at least two vector-like fermions are needed~\cite{Aoki:2011yk}.
}.
With the relevant interactions given by 
\begin{align}
    -\mathcal{L}_{\text{AKY}} &= \lambda_{3/2} (\Phi_{3/2}^\dagger \Phi_1) (\tilde{\Phi}_2^\dagger \Phi_1) + \mathrm{h.c.} \notag \\
    &+ f_{\psi_L} \overline{\ell_L^C} \psi_L \Phi_{3/2} + f_{\psi_R} \overline{\ell_L} \psi_R \Phi_2 + \mathrm{h.c.}, 
\end{align}
the mass formula of the active neutrino is written as
\begin{align}
    m_\nu \sim \frac{f_{\psi_L} f_{\psi_R} \lambda_{3/2} v^2}{(16\pi^2) m_\psi} \mathcal{I}_{\text{AKY}}.
\end{align}
The $\lambda_{3/2}$ coupling should be suppressed by the $\mathrm{SU}(2)_{R_1} \otimes \mathrm{SU}(2)_{R_2} \otimes \mathrm{SU}(2)_{R_3}$ symmetry.
In this model, the lightest $Z_2$-odd particle can be the dark matter candidate.

This class also includes the particle contents of the GNR model: there are $Z_2$-odd $\mathrm{SU}(2)_L$ doublet $\Phi_2$ ($Y=1/2$), $Z_2$-odd singlet $S^\pm$ ($Y=1$), and $Z_2$-even singlet $k^{\pm \pm}$ ($Y=2$) in $\mathbb{M}_{13}$, $\mathbb{M}_{24}$ and $\mathbb{M}_{34}$, respectively. 
The relevant Lagrangian of the GNR model is written by 
\begin{align}
    -\mathcal{L}_{\text{GNR}} &= \frac{\lambda_5}{2} (\Phi_1^\dagger \Phi_2)^2 + \mu \tilde{\Phi}_1^\dagger \Phi_2 S^- + \sigma S^- S^- k^{++} \notag \\
    &+ f_R \overline{e_R^C} e_R k^{++} + \xi \tilde{\Phi}_2^\dagger \Phi_1 S^+ k^{--} + \mathrm{h.c.}.
\end{align}
The neutrino mass is generated via the three-loop diagram as
\begin{align}
    m_\nu \sim \frac{f_R m_{\ell} m_{{\ell^\prime}} \sin 2\beta}{(16 \pi^2)^3 M} \mathcal{I}_{\text{GNR}},
\end{align}
where $\beta$ is the mixing angle between $S^\pm$ and the charged component of $\Phi_2$.
The loop function $\mathcal{I}_{\text{GNR}}$ is approximately proportional to $\lambda_5$ and $\mu$ with a dependence of $\sigma$ and $\xi$.
These couplings should be suppressed, if the flavor symmetry is respected.
The lightest $Z_2$-odd neutral scalars in $\Phi_2$ can be the dark matter candidate.

If the $Z_2$-odd RHNs are additionally introduced in this class, the neutrino masses can be generated by the mechanism proposed in the KNT model.
There are $Z_2$-even singlet $h^\pm$ ($Y=1$) and the $Z_2$-odd singlet $S^\pm$ ($Y=1$) in $\mathbb{M}_{34}$ and $\mathbb{M}_{23}$ (or $\mathbb{M}_{24}$), respectively.
Therefore, by the Lagrangian given by 
\begin{align}
    -\mathcal{L}_{\text{KNT}} = f_L \overline{\ell^C_L} \ell_L h^+ + f_N \overline{N^C} e_R S^+ + \lambda_S (h^+ S^-)^2 + \mathrm{h.c.},
\end{align}
the neutrino masses are calculated as
\begin{align}
    m_\nu \sim \frac{\lambda_S f_L^2 f_N^2 m_\ell m_{\ell^\prime} }{(16 \pi^2)^3 M} \mathcal{I}_{\text{KNT}}.
\end{align}
This model also has the dark matter candidates in the $Z_2$-odd sector.
The $\lambda_S$ coupling can also be suppressed by the flavor symmetry.

\subsubsection*{Class~6: AKS model and N2HDM}

We take $y_3 = y_4 =  1/2$, $\omega_3 = +1$, and $\omega_4 = -1$.
The particle contents in this class contain the $Z_2$-even additional $\mathrm{SU}(2)_L$ doublet $\Phi_2$ in $\mathbb{M}_{13}$ and the $Z_2$-odd singlets with $Y=0$ and $Y=1$ ($\eta$ and $S^\pm$) in $\mathbb{M}_{24}$ or $\mathbb{M}_{34}$.
When the $Z_2$-odd RHNs exist, this class contains the particle contents of the AKS model~\cite{Aoki:2008av,Enomoto:2024jyc}.
Since the softly-broken $Z_2$ symmetry is not imposed here to forbid the FCNC couplings, which was discussed in the original AKS model, the structure of the Yukawa interaction is similar to that in Ref.~\cite{Aoki:2022bkg}.

In the AKS model, the neutrino mass is induced by the following Lagrangian,
\begin{align}
    -\mathcal{L}_{\text{AKS}} = f_N \overline{N^C} e_R S^+ + \kappa \tilde{\Phi}_2^\dagger \Phi_1 S^- \eta + \mathrm{h.c.},
\end{align}
which deduces the neutrino masses at the three-loop level as 
\begin{align}
    m_\nu \sim \frac{\kappa^2 f_N^2 \zeta^2 m_{\ell} m_{\ell^\prime} }{(16 \pi^2)^3 M} \mathcal{I}_{\text{AKS}}.
\end{align}
In this AKS model, the problems of the neutrino mass, the dark matter, and the baryon asymmetry of the Universe can be solved simultaneously by the TeV scale physics.

We note that, since the $Z_2$-even doublet $\Phi_2$ and the $Z_2$-odd singlet $\eta$ are present, this class also encompasses the structure of the N2HDM.
The N2HDM has been extensively studied in the contexts of dark matter and collider phenomenology~\cite{He:2008qm,Grzadkowski:2009iz,Chen:2013jvg,Muhlleitner:2016mzt,Biekotter:2022jyr}.

\begin{table*}[t]
\setlength{\tabcolsep}{6pt}
\centering
\begin{tabular}{|c|c|c|c|c|c|c|}
\hline
 & $(\mathbb{M}_{\Phi_{3/2}},\mathbb{M}_{\Phi_{1/2}},\mathbb{M}_{\Phi_{-1/2}} )^\intercal$ & $(\bm{\chi}^*,\bm{\zeta},\bm{\chi})^\intercal$ & $(S^{++},S^{+},S^0,S^{-},S^{--})^\intercal$ & $\phi_1^\pm$ & $\phi_2^\pm$  \\ \hline
 $\spadesuit$ $Y$ & $3/2, \pm1/2$ & $\pm 1 ,0$ & $\pm 2, \pm 1, 0$ & $0$ & $0$  \\ \hline
 $\spadesuit$ $L$ & $\bm{2}$ & $\bm{3}$ & $\bm{1}$ & $\bm{1}$ & $\bm{1}$  \\ \hline
 $\diamondsuit$ $R_1$ & $\bm{2}$ & $\bm{1}$ & $\bm{1}$ & $\bm{1}$ & $\bm{1}$  \\  \hline
 $\diamondsuit$ $R_2$ & $\bm{3}$ & $\bm{3}$ & $\bm{5}$ & $\bm{1}$ & $\bm{1}$  \\ \hline
\end{tabular}
\caption{
Scalar particle contents below $\Lambda$ in the $n=4$ case for the triplet models (Class 7).
The hypercharges shown here are for $\mathrm{SU}(2)_{R_2}$ components.
}
\label{tab:n=4_2}
\end{table*}

\begin{table*}[t]
\setlength{\tabcolsep}{6pt}
\centering
\begin{tabular}{|c|c|l|l|}
\hline
 Class & $n$ & Charge assignment of the UV model & Higgs sectors as a part of the low-energy effective theory \\ \hline
 1 & $2$ & $y_2 = 1/2$ & xSM \\ \hline
 2 & $3$ & $(y_3,\omega_3)=(1/2,+1)$ & xSM, 2HDM, Zee, CxSM \\ \hline
 3 & $3$ & $(y_3,\omega_3)=(1/2,-1)$ & xSM, IDM (Tao--Ma), SDM \\ \hline
 4 & $3$ & $(y_3,\omega_3)=(3/2,+1)$ & xSM, Zee--Babu \\ \hline
 5 & $4$ & $(y_3,\omega_3,y_4,\omega_4)=(1/2,-1,3/2,-1)$ & xSM, IDM (Tao--Ma), SDM, Zee--Babu, AKY, KNT, GNR \\ \hline
 6 & $4$ & $(y_3,\omega_3,y_4,\omega_4)=(1/2,+1,1/2,-1)$ & xSM, Zee, CxSM, IDM (Tao--Ma), SDM, KNT, AKS, N2HDM \\ \hline
 7 & $4$ & $(T_1,T_2,T_3)^\intercal \sim (\bm{2},\bm{1},\bm{3}),~ T_4 \sim (\bm{1},\bm{2},\bm{1})$ & xSM, 2HDM, Zee, Zee--Babu, GM \\ \hline 
\end{tabular}
\\[0.5cm]
\scalebox{0.81}{
\begin{tabular}{|c|c|c|c|c|c|c|c|c|c|}
\hline
 Models & xSM  & 2HDM & Zee & CxSM & IDM (Tao--Ma) & SDM & Zee-Babu & AKY  \\ \hline
 $(Y,L,Z_2)$ & $(0,\bm{1},+)$ $\in \mathbb{R}$ & $(1/2,\bm{2},+)$ & $(1/2,\bm{2},+)$, $(1,\bm{1},+)$ & $(0,\bm{1},+)$ $\in \mathbb{C}$ & $(1/2,\bm{2},-)$ & $(0,\bm{1},-)$ & $(2,\bm{1},+)$, $(1,\bm{1},+)$ & $(1/2,\bm{2},-)$, $(3/2,\bm{2},-)$ \\ \hline
 Class & 1,2,3,4,5,6,7  & 2,6,7 & 2,6,7 & 2,6 & 3,5,6 & 3,5,6 & 4,5,7 & 5 \\ \hline
\end{tabular}
}
\scalebox{0.88}{
\begin{tabular}{|c|c|c|c|c|c|}
\hline
 Models & KNT & GNR & AKS & N2HDM & GM \\ \hline
 $(Y,L,Z_2)$ & $(1,\bm{1},-)$, $(1,\bm{1},+)$  & $(1/2,\bm{2},-)$, $(1,\bm{1},-)$, $(2,\bm{1},+)$ & $(1/2,\bm{2},+)$, $(0,\bm{1},-)$, $(1,\bm{1},+)$ & $(1/2,\bm{2},+)$, $(0,\bm{1},-)$ & $(1,\bm{3},+)$, $(0,\bm{3},+)$ \\ \hline
 Class & 5,6 & 5 & 6 & 6 & 7 \\ \hline
\end{tabular}
}
\caption{Summary of classification of the extended Higgs models.
}
\label{tab:summary}
\end{table*}

\subsection*{Models for $\mathrm{SU}(2)_L$ triplet scalar}

\subsubsection*{Class~7: GM model}

We consider the degenerated three flavor fermions $(T_1,T_2,T_3)^\intercal$ with the Dirac mass $m$ and one additional flavor $T_4$ with $m_4 \neq m$.
The Majorana masses are taken to be zero, $M_{ij} = 0$.
In this case, the multiplet $(T_1,T_2,T_3)^\intercal$, which is the part of $\bm{T}$, behaves as $\bm{6}$ of $\mathrm{Sp}(6)$.

From the branching rule of $\mathfrak{sp}(6) \supset \mathfrak{su}(2) \oplus \mathfrak{su}(2)$, $\bm{6}$ can be decomposed into $(\bm{2}, \bm{3})$, namely 
\begin{align}
  (T_1,T_2,T_3)^\intercal \sim \bm{2} \otimes \bm{3}.
\end{align}
We regard the former $\mathrm{SU}(2)$ as $\mathrm{SU}(2)_L$ and the latter one as $\mathrm{SU}(2)_{R_2}$.
We also have $T_4$, which is $\bm{2}$ of the other $\mathrm{SU}(2)_{R_1}$.
Therefore, there are the fundamental fermions represented by $(\bm{2},\bm{1},\bm{3})$ and  $(\bm{1},\bm{2},\bm{1})$ under $\mathrm{SU}(2)_L \otimes \mathrm{SU}(2)_{R_1} \otimes \mathrm{SU}(2)_{R_2}$.

By assuming $\mathrm{SU}(2)_L$ and the third component of $\mathrm{SU}(2)_{R_1}$ and $\mathrm{SU}(2)_{R_2}$ are gauged to give the hypercharge, we obtain the following scalar bound states as totally anti-symmetric representations:
\begin{align}
  &(\bm{2},\bm{2},\bm{3}) \sim  (\mathbb{M}_{\Phi_{3/2}},\mathbb{M}_{\Phi_{1/2}},\mathbb{M}_{\Phi_{-1/2}} )^\intercal, \notag \\
  &(\bm{3},\bm{1},\bm{3}) \sim  (\bm{\chi}^*,\bm{\zeta},\bm{\chi})^\intercal, \notag \\
  &(\bm{1},\bm{1},\bm{5}) \sim (S^{++},S^{+},S^0,S^{-},S^{--})^\intercal, \notag \\
  &(\bm{1},\bm{1},\bm{1}) \sim \phi_1^\pm, \phi_2^\pm.
\end{align}
The fields shown here correspond to the components of $\mathrm{SU}(2)_{R_2}$.

We summarize the particle contents in this class in Table~\ref{tab:n=4_2}.
The hypercharges shown here are for $\mathrm{SU}(2)_{R_2}$ components.

The three bi-doublet states $\mathbb{M}_{\Phi_{3/2}},\mathbb{M}_{\Phi_{1/2}}$, and $\mathbb{M}_{\Phi_{-1/2}}$ making $\bm{3}$ of $\mathrm{SU}(2)_{R_2}$ have the hypercharge $Y=3/2,1/2$, and $-1/2$, respectively.
The bi-doublet field $\mathbb{M}_{\Phi_{1/2}}$ (or $\mathbb{M}_{\Phi_{-1/2}}$) corresponds to the SM Higgs doublet.
The bi-triplet field under $\mathrm{SU}(2)_L \otimes \mathrm{SU}(2)_{R_2}$ contains $Y=1$ triplet $\bm{\chi}=(\chi^{++},\chi^+,\chi^0)$ and $Y=0$ triplet $\bm{\zeta} = (\zeta^{+},\zeta^0,\zeta^-)$.
There are the quintuplet field of the $\mathrm{SU}(2)_{R_2}$, which contains doubly, singly, and neutrally charged $\mathrm{SU}(2)_L$ singlet scalars. 
Similar to the previous models, we also have the $Y=0$ parity-even and odd scalars $\phi_1^\pm$ and $\phi_2^\pm$, which are the singlet states under any of the flavor symmetry.

This class corresponds to the GM model, in which $\mathrm{SU}(2)_L$ triplet scalars with $Y=0$ and $Y=1$ are involved.
When the electroweak symmetry breaking occurs as $\langle \mathbb{M}_{\mathrm{SM}} \rangle \propto a \mathbb{I}_2$ and $\langle \chi^0 \rangle = \langle \zeta^0 \rangle \propto b$ satisfying $\sqrt{a^2 + 8b^2} = v$, the rho parameter is unity at the tree level, as in the GM model.

\subsection{Yukawa interaction \label{sec:Yukawa}}

In this subsection, we discuss the Yukawa interaction based on a similar mechanism to the partial compositeness~\cite{Kaplan:1991dc,Agashe:2004rs,Mrazek:2011iu}.
The Yukawa interaction in the low-energy effective theory is provided by the confinement in a part of higher dimensional operators. 
In the UV sector, a massive scalar is introduced to violate the chiral symmetry of the SM quarks and leptons.

Letting $\psi$ be a SM quark or lepton, or a right handed neutrino, we consider a case where the dimension four Yukawa interaction 
\begin{align}
    -\mathcal{L}_Y^{(d = 4)} = \sum_\mathcal{S} \sum_{\phi_\mathcal{S} \in \mathcal{S} } \sum_{(\psi_L,\psi_R)} y^{\phi_\mathcal{S}}_{(\psi_L,\psi_R)} \overline{\psi_{L}} \psi_{R} \phi_\mathcal{S} + \mathrm{h.c.},
    \label{eq:Yukawa1}
\end{align}
is deduced in an extended Higgs model.
We denote $\mathcal{S}$ as a set of $\mathrm{SU}(2)_H$ singlet scalar bosons $\phi_\mathcal{S}$, all of which have the same quantum charges under the SM gauge group.
The Yukawa matrices, in which a combination of $\psi_L$, $\psi_R$, and $\phi_\mathcal{S}$ are involved, are represented by $y^{\phi_\mathcal{S}}_{(\psi_L,\psi_R)}$.
The flavor indices of $\psi$ are omitted.

To obtain the Yukawa interaction, we introduce a scalar field $\Phi_\mathcal{S}$ with a mass $M_{\Phi_\mathcal{S}} (\gtrsim \Lambda)$, which has the quantum charges of $\phi_\mathcal{S}$.
By denoting $(TT)_{\phi_\mathcal{S}}$ as a $\mathrm{SU}(2)_H$ singlet state reproducing $\phi_\mathcal{S}$ after the confinement, namely $(TT)_{\phi_\mathcal{S}} \sim \Lambda^2 \phi_\mathcal{S}$, we have the following UV Lagrangian:
\begin{align}
  -\mathcal{L}_{\mathrm{UV}} &= \sum_\mathcal{S}  \sum_{(\psi_L,\psi_R)} \mathcal{Y}_{(\psi_L,\psi_R)} \overline{\psi_{L}} \psi_{R} \Phi_\mathcal{S} + \mathrm{h.c.} \notag \\
  &+ \sum_\mathcal{S} \sum_{\phi_\mathcal{S} \in \mathcal{S} } \mathcal{H}^{\phi_\mathcal{S}} (TT)_{\phi_\mathcal{S}} \Phi_\mathcal{S}^\dagger + \mathrm{h.c.}.
\end{align}
By integrating out $\Phi_\mathcal{S}$, we effectively obtain the four-fermi operators as written by
\begin{align}
    -\mathcal{L}_{\mathrm{eff}} = \sum_\mathcal{S} \sum_{\phi_\mathcal{S} \in \mathcal{S} }  \sum_{(\psi_L,\psi_R)} \frac{\mathcal{H}^{\phi_\mathcal{S}} \mathcal{Y}_{(\psi_L,\psi_R)} } {M^2_{\Phi_\mathcal{S}}} \overline{\psi_{L}} \psi_{R} (TT)_{\phi_\mathcal{S}}.
\end{align}
After the confinement, we obtain the Yukawa interaction as Eq.~\eqref{eq:Yukawa1}, namely $y^{\phi_\mathcal{S}}_{(\psi_L,\psi_R)} \sim \mathcal{H}^{\phi_\mathcal{S}} \mathcal{Y}_{(\psi_L,\psi_R)} \Lambda^2 /M_{\Phi_\mathcal{S}}^2$.

Since only $\mathcal{Y}_{(\psi_L,\psi_R)}$ has the flavor indices of $\psi$, the multiple Yukawa matrices $y^{\phi_\mathcal{S}}_{(\psi_L,\psi_R)}$ have the same flavor structure.
For example, in the 2HDM, the Yukawa interaction of the up-type quarks ($u,c,t$) is obtained as
\begin{align}
    -\mathcal{L}_{Y,\mathrm{2HDM}}^{(d=4)} &= \mathrm{diag}(y_u,y_c,y_t) \overline{Q_{L}} u_{R} \Big( \tilde{\Phi}_1 + \zeta \tilde{\Phi}_2 \Big),
\end{align}
where $Q_L$ ($u_R$) is the left-handed (right-handed) SM quark doublet (singlet), and $\zeta = \mathcal{H}^{\Phi_2} / \mathcal{H}^{\Phi_1}$.
The additional Yukawa interactions for the down-type quarks and the leptons are also controlled by the same factor $\zeta$.
Therefore, in our setup, the minimal flavor violating Yukawa interaction (so-called Yukawa alignment~\cite{Pich:2009sp}) is realized. 
The Yukawa interactions for different models, such as a model with a charged $\mathrm{SU}(2)_L$ singlet, can also be obtained by introducing other heavy scalars $\Phi_{\mathcal{S}}$.

\section{Discussions \label{sec:Discussions}}

In Table.~\ref{tab:summary}, we summarize the list of the extended Higgs models as a part of the low-energy effective theory, which are classified by the flavor number and the charge assignment of the $\mathrm{SU}(2)_H$ theory.

In this paper, we discussed the relation between the scalar particle contents in the low-energy effective theory and the flavor number and the charge assignment in the UV gauge theory, assuming implicitly the existence of a specific energy scale like the Landau pole, where the low-energy theory is replaced by the UV theory.
In this sense, the deduced extended Higgs models should be regarded as valid only for energy scales below the confinement scale $\Lambda$.
Moreover, the Landau-pole scale, which is estimated in the literature for the 2HDM~\cite{Kominis:1993zc,Kanemura:1999xf,Chowdhury:2015yja}, IDM~\cite{Goudelis:2013uca,Kanemura:2023wap,Bandyopadhyay:2025ilx}, GM model~\cite{Blasi:2017xmc}, etc., is the result of extrapolating the low-energy theory towards the UV.
We expect that such apparent pathological behavior of the IR effective theory around that scale is cured once the description is replaced by the asymptotically free $\mathrm{SU}(2)_H$ gauge theory.
However, whether the Landau pole really exists or not strongly depends on the parameters of the model.
It is non-trivial whether our discussion can be applied to the case where the Landau pole does not exist below the Planck scale.
As it is strongly parameter-dependent, we do not discuss this issue in this paper.

In this paper, we have not pursued the question of how to reproduce the 125~GeV Higgs boson.
The mass spectrum of the bound states is expected to be linked to the non-perturbative dynamics of the $\mathrm{SU}(2)_H$ gauge theory at the scale $\Lambda$, 
which lies beyond the scope of our analysis.
We briefly mention several possibilities for the mechanism to generate a light Higgs boson in the context of strong gauge dynamics.
One possibility is the pseudo-Nambu--Goldstone boson (pNGB) scenario, in which the Higgs boson arises as a pNGB associated with the spontaneous breaking of a global symmetry, 
such as 
$\mathrm{SU}(4) \to \mathrm{Sp}(4) \supset \mathrm{SU}(2) \otimes \mathrm{SU}(2)$ in Class~1~\cite{Lewis:2011zb,Cacciapaglia:2014uja,Arbey:2015exa,Arthur:2016dir,Arthur:2016ozw,Drach:2017btk}, triggered by the confinement of the underlying gauge theory.
Another possibility is due to the renormalization group running of the Higgs mass parameters below the confinement scale $\Lambda$~\cite{Inoue:1982ej,Inoue:1982pi}.
Even if only small mass differences among the bound states are generated at the scale $\Lambda$, the differences can be enhanced by the running due to the steep change 
of the mass parameters and the quartic couplings at around $\Lambda$.
While the present study does not commit to any specific mechanism, investigating the mechanism to reproduce the 125~GeV Higgs boson should be an important future work.

Connecting the UV theory and the low-energy effective theory has been done in the context of supersymmetric (SUSY) models~\cite{Harnik:2003rs,Kanemura:2012uy}, in which the low-energy effective potential can be written down by using the UV/IR duality~\cite{Intriligator:1995au}.
A SUSY gauge theory with $n$ flavors is described by the $n$ pairs of chiral superfields, $(\hat{T}_{Li}, \hat{T}_{Ri}^c)=(\tilde{T}_{Li}+\sqrt{2}\theta\chi_i+\theta^2 F_{Li}, \tilde{T}_{Ri}^* +\sqrt{2}\theta\eta_i^*+\theta^2 F_{Ri}^*)$.
Here, the bound states $\hat{M}_{LR}^{ij}\sim \hat{T}_{Li}\hat{T}_{Rj}^c$ ($\hat{M}^{ij}_{LL}\sim \hat{T}_{Li}\hat{T}_{Lj}$) is
an independent super field of $\hat{M}^{ji}_{LR}\sim \hat{T}_{Ri}^c\hat{T}_{Lj}$ ($\hat{M}^{ji}_{RR}\sim \hat{T}_{Ri}^c\hat{T}_{Rj}^c$),
in contrast to the non-SUSY case with $n$ fundamental Dirac fermions, 
where the bound state $M^{ij}_{\eta\chi}\sim\eta_i^*\chi_j$ ($M^{ij}\sim \eta_i\eta_j$) is related to
$M^{ji}_{\eta\chi}\sim\chi_i\eta_j^*$ 
($M^{ij}_{\eta\eta}\sim \eta_i^*\eta_j^*$) by charge conjugation.
Thus, the number of scalar degrees of freedom below $\Lambda_H$ in the SUSY gauge theory
is doubled from that of the non-SUSY one.
It is actually shown in the comparison between
the SUSY $\mathrm{SU}(2)_H$ theory with $n=3$ studied in Refs.~\cite{Harnik:2003rs,Kanemura:2012uy} 
and the model discussed in Sec.~\ref{sec:n=3case}.

In Sec.~\ref{sec:Yukawa}, we discussed the Yukawa interactions in our framework and presented a natural realization of Yukawa alignment in the 2HDM at the scale $\Lambda$ as an example.
However, we do not address the origin of the quark and lepton flavor structures of the SM.
Although this question is beyond the scope of this paper, there exist various mechanisms in the literature that aim to explain the observed flavor hierarchies and mixings (e.g., introducing flavor symmetries~\cite{Froggatt:1978nt,Ishimori:2010au,Altarelli:2010gt}, invoking strong dynamics~\cite{Strassler:1995ia,Nelson:1996km,Hamada:2022ino}, radiatively generating fermion masses~\cite{Balakrishna:1987qd,Balakrishna:1988ks}, and so on).
We note that additional flavor mixing will be generated by the renormalization group running below $\Lambda$. 
In addition, the extra scalar contribution, such as a charged Higgs boson exchange, can provide deviations from the SM predictions for flavor observables, even if the minimal flavor violation structure is kept at the electroweak scale.

In this paper, we have considered an $\mathrm{SU}(2)_H$ gauge theory as a UV completion of various extended Higgs models.
However, it is non-trivial whether this picture can be consistently maintained all the way up to the Planck scale.
Above the confinement scale, the gauge group is 
$G_{\mathrm{SM}} \otimes \mathrm{SU}(2)_H$, 
and one may consider a grand unification of these gauge couplings at some higher scale.
Representative examples of unified gauge groups embedding the SM gauge groups are 
$G_{\mathrm{GUT}} = \mathrm{SU}(5)$~\cite{Georgi:1974sy}, $\mathrm{SO}(10)$~\cite{Georgi:1974my,Fritzsch:1974nn}, etc., 
and a straightforward extension is to consider a GUT based on $G_{\mathrm{GUT}} \otimes \mathrm{SU}(2)_H$.
In such a framework, the vector-like fundamental fermions of $\mathrm{SU}(2)_H$ can be charged under $G_{\mathrm{GUT}}$ without introducing additional gauge anomalies.
One may also consider embedding $G_{\mathrm{SM}} \otimes \mathrm{SU}(2)_H$ into a larger group in which $\mathrm{SU}(2)_H$ appears as a subgroup.
However, a fully realistic implementation of gauge coupling unification involving $\mathrm{SU}(2)_H$ is left for future work~\cite{Futurework}.

Finally, in this paper, a specific shape of the extended Higgs models protected by symmetries are naturally derived.
To obtain more general Higgs potentials, it is necessary to take into account the breaking term of the flavor symmetries.
For example, in the case of the 2HDM, the exact flavor symmetry forbids $\lambda_4$, $\lambda_5$, $\lambda_6$, $\lambda_7$ and $\mu_3$.
These parameters emerge as effects of the symmetry breaking.
Naively, all these couplings are expected to appear at the similar order.
However, this discussion holds at the confinement scale and can change at low energy due to various effects.
It is not obvious whether the experimental constraints on the 2HDM are properly satisfied by such a more general 2HDMs.
At this stage, it is difficult to conclude whether our framework can truly reproduce various existing Higgs models at the low energy.
Regardless of the question whether the Higgs models predicted by our framework at the low energy coincide with the existing ones or not, it is an extremely interesting question whether our Higgs models deduced from the UV theory can explain the problems beyond the SM.
In this paper, we do not pursue this issue further, and this will be examined elsewhere~\cite{Futurework}.

\section{Conclusions \label{sec:Conclusions}}

In this paper, we have considered the $\mathrm{SU}(2)_H$ gauge theory with confinement as a UV theory of the extended Higgs sectors.
We have investigated the relation between the scalar particle contents at the low energy and flavor number and the charge assignment of the $\mathrm{SU}(2)_H$ theory.
We have found that particle contents of various extended Higgs sectors previously proposed to explain the beyond the SM problems are deduced by the $\mathrm{SU}(2)_H$ theory, as a part of the low-energy theory.
Our findings may provide a new picture for the UV completion of the extended Higgs sectors.

\begin{acknowledgments}
SK was supported, in part, 
by Grants-in-Aid for Scientific Research(KAKENHI) Nos.~24KF0060, 24KF0238, and 23K17691.
SK and TS were supported, in part, by 
Grants-in-Aid for Scientific Research(KAKENHI) No.~20H00160.
YM was supported by the JSPS Grant-in-Aid for JSPS Fellows No. 23KJ1460 and JSPS KAKENHI No.~22K21347.
\end{acknowledgments}

\appendix

\bibliography{ref}

\end{document}